\documentclass[preprint,showpacs,amsmath,amssymb]{revtex4}

\usepackage{graphicx,amsmath,amssymb,bm,multirow}
\usepackage{amsthm}
\usepackage{amsfonts}
\usepackage{dcolumn}
\usepackage{bm}
\usepackage[usenames,dvipsnames]{color}

\begin{document}
 \preprint{preprint}

\title{Simultaneous quadrupole and octupole shape phase transitions in Thorium}
\author{Z. P. Li}
\affiliation{School of Physical Science and Technology, Southwest University, Chongqing 400715, China}
\author{B. Y. Song}
\affiliation{School of Physical Science and Technology, Southwest University, Chongqing 400715, China}
\author{J. M. Yao}
\affiliation{School of Physical Science and Technology, Southwest University, Chongqing 400715, China}
\author{D. Vretenar}
\affiliation{Physics Department, Faculty of Science, University of Zagreb, 10000 Zagreb, Croatia}
\affiliation{Kavli Institute for Theoretical Physics China, CAS, Beijing 100190, China}
\author{J. Meng}\thanks{mengj@pku.edu.cn}
\affiliation{State Key Laboratory of Nuclear Physics and Technology, School of Physics,
Peking University, Beijing 100871, China}
\affiliation{School of Physics and Nuclear Energy Engineering, Beihang University, Beijing 100191, China}
\affiliation{Department of Physics, University of Stellenbosch, Stellenbosch, South Africa}

\begin{abstract}
The evolution of quadrupole and octupole shapes in Th isotopes is studied in the framework of nuclear Density Functional Theory.
Constrained energy maps and observables calculated with microscopic
collective Hamiltonians indicate the occurrence of a simultaneous quantum
shape phase transition between spherical and quadrupole-deformed prolate shapes,
and between non-octupole and octupole-deformed  shapes, as functions of the neutron
number.  The nucleus $^{224}$Th is closest to the critical point of a double
phase transition. A microscopic mechanism of this phenomenon is discussed in
terms of the evolution of single-nucleon orbitals with deformation.
\end{abstract}

\pacs{21.60.Jz, 21.60.Ev, 21.10.Re, 21.10.Tg}
\maketitle



Quantum phase transitions (QPTs) present a very active field of research in
condensed matter as well as in mesoscopic systems: atomic nuclei,
molecules, and atomic clusters.  Nuclear QPTs, in particular, correspond
to shape transitions  between competing ground-state phases induced by variation of a
non-thermal control parameter (number of nucleons) at zero temperature
\cite{Iac.03,Rick.06,Rick.07,CJ.09,CJ.10}. Most experimental and theoretical studies
of first- and second-order nuclear QPT have considered quadrupole collective
degrees of freedom, either for axially-symmetric deformed
shapes~\cite{FI.00,CZ.00,FI.01,CZ.01,Pietralla04,Meng05} or triaxial shapes~\cite{Jolie01,FI03}.
Much less analyzed, although
potentially very interesting, are transitions related to stable or dynamical
octupole shapes (reflection-asymmetric, pear-like shapes).
Phenomenological geometric models
of nuclear shapes and potentials~\cite{Bona05,Lenis06,Bizz04,Bizz08,Jol12},
algebraic models of nuclear structure~\cite{Kuyu02} and, more recently,
microscopic energy density functionals~\cite{Zhang10,Guo10} have been
employed in studies of this type of transitions. In this Letter
we report the first microscopic study of a simultaneous double QPT
between spherical and quadrupole-deformed prolate shapes,
and between non-octupole and octupole-deformed shapes in Th isotopes,
located close to $^{224}$Th.

Studies of shape transitions usually start with the calculation of potential energy surfaces (PESs)
as functions of collective deformation variables. The microscopic analysis of
quadrupole and octupole shapes presented in this Letter is based on nuclear
covariant density functional theory (CDFT)~\cite{Ring96,Vret05,Meng06}, which
has successfully been applied to the description of a variety of structure phenomena
over the entire chart of nuclides.

In the present analysis the relativistic mean-field (RMF)
implementation of the CDFT framework is employed, using the functional PC-PK1~\cite{Zhao10} and
with pairing correlations treated in the BCS approximation.
The microscopic PESs are obtained by performing constrained RMF+BCS calculations, with
constraints on both quadrupole and octupole mass moments~\cite{Geng07,Lu12}.
To study the occurrence of possible phase transitions,
the behavior of observables that can be related to
order parameters (equilibrium deformations, ground-state charge radii,
excitation energies, electromagnetic transition rates, etc.) must be analyzed.
Observables that characterize low-lying collective excitations associated with
quadrupole and octupole degrees of freedom are determined by the
eigenstates of the corresponding generalized collective Hamiltonian (CH),
with deformations as dynamical collective coordinates.
The self-consistent solutions of constrained RMF+BCS calculations:
single-particle wave functions, occupation probabilities, and quasiparticle energies
that correspond to each point on the energy surface, are used to calculate
the parameters that determine the dynamics of the CH:  the collective potential and
inertia parameters as functions of collective coordinates.
In the simplest approximation the moments of inertia are calculated using the Inglis-Belyaev
formula \cite{Inglis56,Belyaev61},
the mass parameters associated with the two quadrupole collective coordinates are computed in the cranking
approximation \cite{GG.79,Lib.99,RR.12},
and the collective potential is obtained by subtracting the energy of the zero-point motion from
the total-energy surface \cite{GG.79}. The detailed expressions are given in Ref.~\cite{Li09}.
The diagonalization of the Hamiltonian yields the excitation spectra and collective
wave functions that are used in the calculation of various observables, e.g., electromagnetic transition
probabilities \cite{Niksic09,NVR.11,Guzm12}.

\begin{figure*}[htb!]
\centering
\includegraphics[scale=0.35]{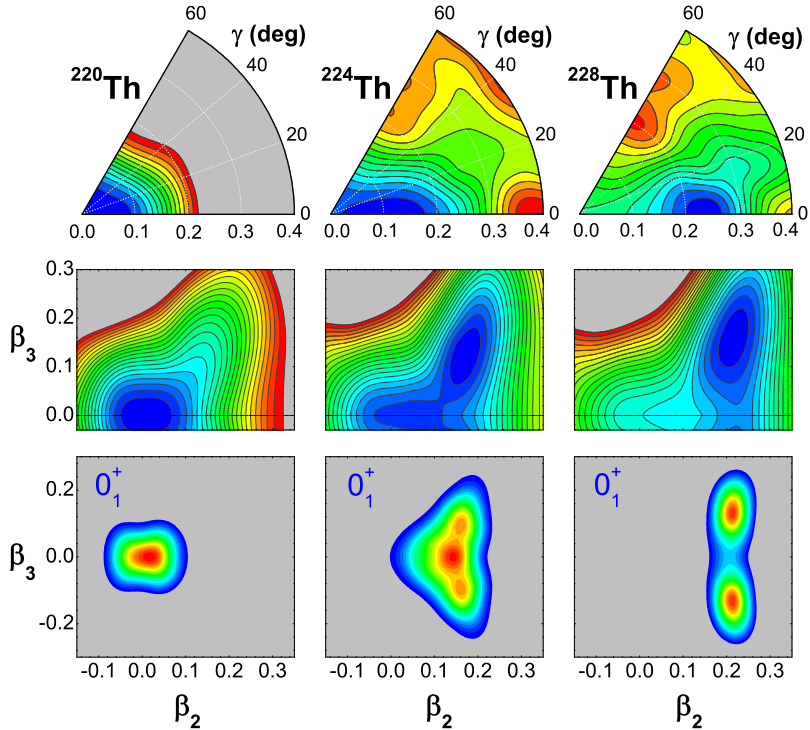}
\caption{\label{fig:PES2}(Color online)  Self-consistent RMF+BCS triaxial quadrupole energy surfaces
in the $\beta_2$-$\gamma$ plane ($0\le \gamma\le 60^0$) (upper panels),
and axially-symmetric quadrupole-octupole energy surfaces in the $\beta_2$-$\beta_3$ plane
(middle panels), for $^{220,224,228}$Th. The contours join points on the surface with the same
energy, and the separation between neighboring contours is 0.5~MeV.
Probability density distributions for the ground states $0^+_1$
of $^{220,224,228}$Th in the $\beta_2$-$\beta_3$ plane (lower panels).}
\end{figure*}

The self-consistent energy surfaces of $^{220,224,228}$Th, calculated in the
constrained RMF+BCS model with the PC-PK1 density functional
in the particle-hole channel and a separable
pairing force~\cite{TMR09} in the particle-particle channel, are plotted in the
panels of the upper two rows of Fig.~\ref{fig:PES2}. The contours join points on
the surface with the same energy; successive contours differ in energy by 0.5 MeV.
The upper panels display triaxial quadrupole PESs
in the $\beta_2$-$\gamma$ plane. One notices the rather rapid transition from the
spherical $^{220}$Th to the well-deformed prolate shape of $^{228}$Th. An
interesting feature of the isotopic evolution
is the flat prolate minimum in $^{224}$Th, that extends in the interval $0 \leq \beta \leq 0.2$
of the axial deformation parameter. Flat-bottom potentials are characterized by fluctuations in the
deformation parameters and, therefore, indicate a possible phase transition. The quadrupole
PES of $^{224}$Th is reminiscent of those that we analyzed in the $N \approx 90$
region \cite{Ni07,Li09}, in connection with first-order shape phase transitions between
spherical and prolate axial shapes, first studied by Iachello using the square-well
X(5) model~\cite{FI.01}.

If one considers $^{224}$Th as a system at the X(5) critical point of a first-order
shape phase transition then, noticing that the PES is rather steep in $\gamma$,
the dependence of the PES on the two deformation parameters can be decoupled
in accordance with the assumption of the X(5) model and, instead, the axially-symmetric
PES  in the $\beta_2$-$\beta_3$ plane analyzed. This is shown in the middle
row of Fig.~\ref{fig:PES2}, where we plot the PESs of $^{220,224,228}$Th calculated
using the RMF+BCS model with constraints on the expectation values
of the quadrupole moment $\langle Q_{20}\rangle$, and octupole
moment $\langle Q_{30}\rangle$. The region of actinides with $N\approx134$
is characterized by pronounced octupole correlations \cite{Butler96} and,
indeed, our calculation indicates a rapid shape transition between $N=130$ and $N=138$,
from non-octupole to pronounced octupole deformations.
The PES of the transitional nucleus $^{224}$Th is rather soft with respect to
the octupole deformation in the region $\beta_2\sim 0.16$.
This is consistent with the fact that in the Th isotopic chain the
lowest excitation energy of the $3^-$ state is found in $^{224}$Th~\cite{NNDC}.
The rapid transition and critical behavior is also clearly illustrated
by the probability density distributions for the ground
states $0^+_1$ of $^{220,224,228}$Th in the $\beta_2$-$\beta_3$ plane,
plotted in the lower panels of Fig.~\ref{fig:PES2}.
The $0^+_1$ states are eigenstates of the
axially-symmetric quadrupole-octupole vibrational collective Hamiltonian (2D-CH),
with parameters determined by the self-consistent constrained RMF+BCS calculations.
The probability densities for $^{220}$Th and $^{228}$Th isotopes
peak at the spherical and stable quadrupole-octupole deformed shapes, respectively,
whereas for $^{224}$Th it is extended along both $\beta_2$
and $\beta_3$.


\begin{figure}[tb!]
\centering
\includegraphics[width=8.5cm]{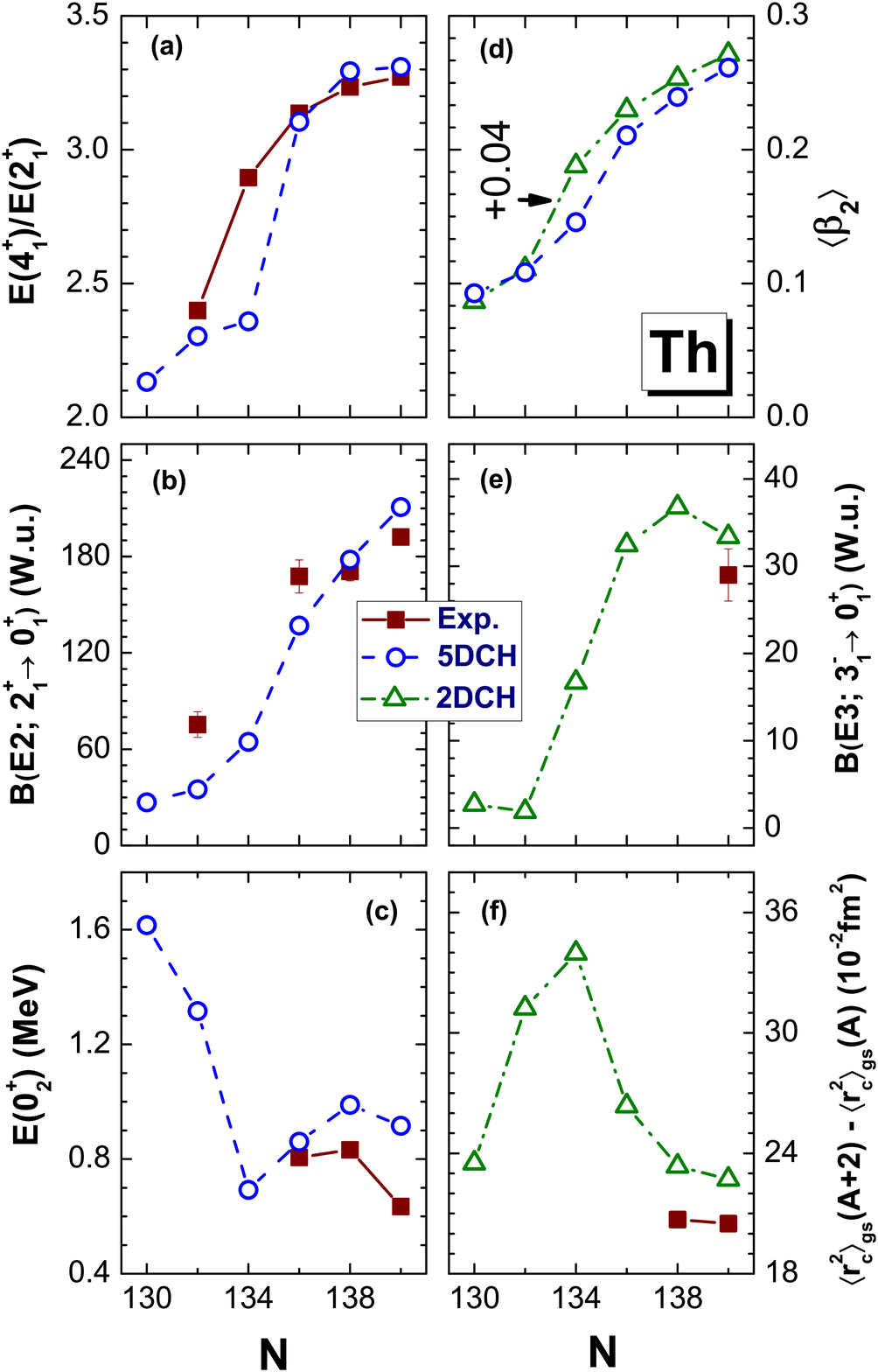}
\caption{\label{fig:obs}(Color online) Evolution of the energy ratios $E(4^+_1)/E(2^+_1)$ (a),
$B(E2; 2^+_1\to 0^+_1)$ values (b),
excitation energies of $0^+_2$ states (c), expectation values of the quadrupole deformation
parameter $\langle\beta_2\rangle$ in the ground state $0^+_1$ (d),
$B(E3; 3^-_1\to 0^+_1)$ (e),
and the isotope shifts of the ground-state charge radii:
$\langle r^2_c\rangle_{\rm gs}(A+2)-\langle r^2_c\rangle_{\rm gs}(A)$ (f),
as functions of the neutron number in Th isotopes.
Microscopic values calculated with the 5D (three-dimensional rotations and
$\beta_2$-$\gamma$ quadrupole vibrations) (circles), and 2D (axially-symmetric
quadrupole-octupole vibrations) (triangles) collective Hamiltonians
based on the PC-PK1 density functional are compared to available data.}
\end{figure}

To find quantitative signatures of possible shape QPTs, we investigate the
dependence of observables that can be related to order parameters as
functions of the control parameter -- nucleon number. A critical point of a QPT is
characterized by a sudden change in the order parameter, even though
one expects that in systems with a finite number of particles the transition is,
to a certain extent, smoothed out. In Figure~\ref{fig:obs} we analyze the evolution
with neutron number of several quantities that are directly computed using the
excitation energies and collective wave functions obtained with the 5D
collective Hamiltonian (CH) (three-dimensional rotations and $\beta_2$-$\gamma$ quadrupole vibrations)
and the 2D-CH (axially-symmetric quadrupole-octupole vibrations).
The parameters of the two Hamiltonians are completely determined by the
self-consistent solutions of RMF+BCS equations with the functional PC-PK1
(cf. Fig.~\ref{fig:PES2}). Figure~\ref{fig:obs} displays the isotopic dependence of the
energy ratios $E(4^+_1)/E(2^+_1)$,
$B(E2; 2^+_1\to 0^+_1)$ values, excitation energies of $0^+_2$ states,
the expected quadrupole deformation $\langle\beta_2\rangle$ of the ground state $0^+_1$,
$B(E3; 3^-_1\to 0^+_1)$ values,
and the isotope shifts of the ground state charge radii:
$\langle r^2_c\rangle_{\rm gs}(A+2)-\langle r^2_c\rangle_{\rm gs}(A)$,
for a sequence of Th nuclei. The theoretical $B(E3; 3^-_1\to 0^+_1)$ value is obtained using the collective wave functions of $K^\pi=0^-$ and $K^\pi=0^+$ states from the 2D-CH calculation. In general, the agreement between the theoretical results and the data is very good, especially considering that the calculation does not involve any additional parameter beside those used in the calculation of the self-consistent PESs.

An important result shown in Fig.~\ref{fig:obs} is that all the considered quantities present pronounced discontinuities at $^{224}$Th. The panels on the left point to a phase transition between spherical and quadrupole-deformed prolate shapes, whereas the right panels reveal a phase transition from non-octupole to octupole-deformed shapes. For both shape QPTs $^{224}$Th appears to be closest to the critical point. An exception is the energy ratio $E(4^+_1)/E(2^+_1)$
in $^{224}$Th, which experimentally is close to the value of $2.9$ predicted by the X(5) model~\cite{FI.01} of the quadrupole QPT,
whereas a much smaller value is calculated by the quadrupole 5D-CH. However, as shown in panel (d), this can simply be due to missing
octupole correlations in the 5DCH calculation. Fig.~\ref{fig:obs}(d) compares the expectation
values of the  quadrupole deformation parameter $\langle\beta_2\rangle$,
calculated with the 5D-CH and 2D-CH. We notice that, in particular for $^{224}$Th,
the quadrupole-octupole Hamiltonian predicts a considerably larger value
of $\langle\beta_2\rangle$, which would then translate in a larger value of the
ratio $E(4^+_1)/E(2^+_1)$.

The quantities determined by the evolution of quadrupole-octupole
correlations (panels on the right of Fig.~\ref{fig:obs}), display a
sudden change at $^{224}$Th and then appear to saturate for
heavier Th isotopes. This is consistent with the systematics of experimental excitation energies
of the states $1_1^-$ and $3_1^-$, which display a parabolic dependence on the mass
number with a minimum between $^{224}$Th and $^{226}$Th. The two-neutron separation energies
also show a small discontinuity at $^{224}$Th, indicating a structural change. We have performed calculations up to $A=234$ and found that the PESs in heavier nuclei become extended and soft with respect to $\beta_3$, but with almost constant $\beta_2\sim 0.25 - 0.3$. Consequently, the excitation energy of the first $K^\pi=0^-$ vibrational state
predicted by the 2D-CH increases with mass number.
This is consistent with the increase of the excitation energy of the experimental
$1_1^-$ state (bandhead of $K^\pi=0^-$ band), which has been interpreted
as a signature of the transition from stable octupole deformations to octupole
vibrations \cite{Bona05,Lenis06,Bizz04,Bizz08}.

\begin{figure}[tb!]
\centering
\includegraphics[scale=0.22]{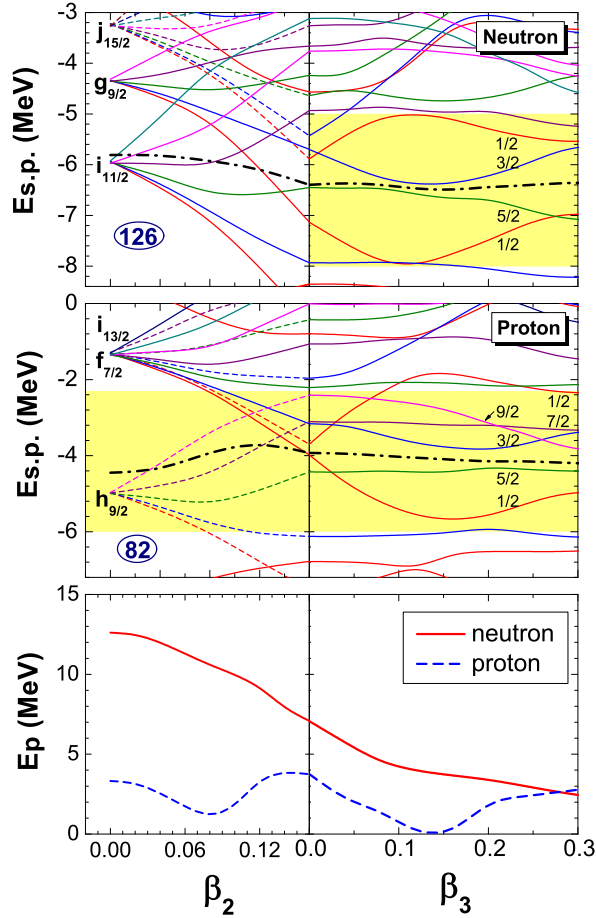}
\caption{\label{fig:Esp} (Color online)
Single-neutron levels (upper panel), single-proton levels (middle panel),
and neutron and proton pairing energies (lower panel) of $^{224}$Th,
as functions of the deformation parameters. Each plot follows
the quadrupole deformation parameter $\beta_2$ up to the position
of the equilibrium minimum $\beta_2 = 0.16$, with the constant octupole
deformation parameter $\beta_3 = 0$ (left panels).
For the constant value $\beta_2 = 0.16$, the panels on the right
display the dependence of the single-nucleon energies and
pairing energies on the octupole deformation, from
$\beta_3 = 0$ to $\beta_3 = 0.3$. The thick dash-dotted (black) curves denote
the Fermi levels.}
\end{figure}

A microscopic picture of the softness of the potential
with respect to both $\beta_2$ and $\beta_3$, and the related phenomenon
of QPTs in $^{224}$Th, emerges from the dependence of the single-nucleon
levels on the two deformation parameters.
In the upper and middle panels of Fig.~\ref{fig:Esp}
we plot the single neutron and proton levels
of $^{224}$Th along a path in the $\beta_2$-$\beta_3$ plane.
The path follows the quadrupole deformation parameter $\beta_2$
up to the position of the equilibrium minimum $\beta_2 = 0.16$, with the
octupole deformation parameter kept constant at zero value. Then,
for the constant value $\beta_2 = 0.16$, the path continues from
$\beta_3 = 0$ to $\beta_3 = 0.3$. In the lower panel of Fig.~\ref{fig:Esp}
we also show the evolution of the neutron and proton pairing energies,
which reflect the level density around the Fermi level (dash-dotted
curves).

In the mean-field approach there is an close relation between the total binding energy and the level density around the Fermi level in the Nilsson diagram of single-particle energies. A lower-than-average density of single-particle levels around the Fermi energy results in extra binding, whereas a larger-than-average value reduces binding. Therefore, the extended quadrupole minimum in $^{224}$Th (upper panel of Fig.~\ref{fig:PES2}) can be mainly  attributed to the wide region of low proton-level density around the Fermi surface, similar to the density of neutron levels in the rare-earth region at $N\approx 90$ for the
X(5) QPT \cite{Ni07,Li09}. In the present case the Fermi surface for
$Z \sim 90$ is far above the $Z=82$ shell closure, but for small quadrupole deformations
the splitting of the intruder state is not large enough to increase the density of states.
On the other hand, the single-neutron level density is determined by the
$N=126$ closure in lighter Th isotopes, and by the splitting of the intruder
state $j_{15/2}$ in the heavier isotopes. Therefore, the lighter and heavier isotopes display nearly spherical
and well-deformed prolate shapes, respectively.
Moreover, the mechanism responsible for the appearance of octupole deformations can be attributed to a parity-breaking odd-multipolarity interaction that couples intrinsic states of opposite parity, as discussed in Ref.~\cite{Butler96}. The necessary condition for the presence of low-energy octupole collectivity is the existence of pairs of orbitals near the Fermi level that are strongly coupled by the octupole interaction. This condition is fulfilled in
$^{224}$Th, for which in Fig.~\ref{fig:Esp} one notices states of opposite parity close to the Fermi level, that originate from the spherical levels $g_{9/2}$ and $j_{15/2}$ for neutrons, and $f_{7/2}$ and $i_{13/2}$ for protons. The octupole minimum, soft along the $\beta_3$-path in $^{224}$Th, is attributed to the low level density of both proton and neutron states close to the corresponding Fermi surface.

%
%

In conclusion, we have analyzed the evolution of quadrupole and octupole shapes in Th isotopes
using a consistent microscopic framework based on CDFT and the generalized collective Hamiltonian, without any phenomenological parameter. The RMF+BCS calculations of constrained quadrupole triaxial and axially-symmetric quadrupole-octupole energy maps have been performed.
The constrained RMF+BCS solutions determine the parameters of a quadrupole 5D, and a quadrupole-octupole 2D collective
Hamiltonians, that are used to calculate observables related to order parameters.
Both the calculated PESs and the predicted observables (excitation energies, isotope shifts of charge radii, electromagnetic
transition rates) indicate the occurrence of a simultaneous phase
transition between spherical and quadrupole-deformed prolate shapes,
and between non-octupole and octupole-deformed  shapes, with $^{224}$Th being
closest to the critical point of the double QPT. A microscopic interpretation
of the QPT has been given in terms of the evolution of single-nucleon
orbitals with quadrupole and octupole deformation parameters. The results
are in good agreement with experiment and with previous studies of
shape transitions in this region. This is, however, the first study in which both
mean-field PESs and order parameters have been calculated microscopically, without any phenomenological
control parameter. Double shape QPT and related critical-point phenomena
could also occur in other finite quantum systems characterized by quadrupole
and octupole collective degrees of freedom.

\bigskip
We thank B. N. Lu for helpful discussions. This work was supported in part by the Major State 973 Program 2013CB834400,
the NSFC under Grant Nos. 11335002, 11175002, 11105110, and 11105111,
the Research Fund for the Doctoral Program of Higher Education under Grant
No. 20110001110087, the Natural Science Foundation of Chongqing cstc2011jjA0376,
and the Fundamental Research Funds for the Central Universities
(XDJK2010B007 and DJK2011B002).


\end{document}